\documentclass[twocolumn,showpacs,preprintnumbers,amsmath,amssymb]{revtex4}

\usepackage{graphicx}
\usepackage{dcolumn}
\usepackage{bm}

\begin{document}


\title{Efficient Non-Resonant Microwave Absorption in Thin Cylindrical Targets: Experimental Evidence}

\author{A. Akhmeteli}
\email{akhmeteli@ltasolid.com}
 \homepage{http://www.akhmeteli.org}
\affiliation{%
LTASolid Inc.\\
10616 Meadowglen Ln 2708, Houston, TX 77042, USA
}%
\author{N. G. Kokodiy}%
\email{kokodiy.n.g@gmail.com}
\author{B.V. Safronov}%
\author{V.P. Balkashin}%
\author{I.A. Priz}%
\affiliation{%
Kharkov National University\\
Ukraine, Kharkov, Svobody sqw., 4
}%
\author{A. Tarasevitch}%
\affiliation{%
University of Duisburg-Essen, Institute of Experimental Physics\\
Lotharstr. 1, 47048 Duisburg, Germany
}%

\date{\today}

\begin{abstract}
Significant (up to 6\%) absorption of microwave power focused on a thin fiber (the diameter is three orders of magnitude less than the wavelength) by an ellipsoidal reflector is demonstrated experimentally. This new physical effect can be used in numerous applications, for example, for efficient heating of nanotubes by laser beams.
\end{abstract}

\maketitle

\section{\label{sec:level1}Introduction}

A theoretical possibility of non-resonant, fast, and efficient heating of extremely thin conducting cylindrical targets by broad electromagnetic beams was described in Ref.~\cite{Akhm4} (see also the sections on the "transverse geometry" in Refs.~\cite{Akhm10,Akhm111} and references there). The diameter of the cylinder can be orders of magnitude smaller than the wavelength of the electromagnetic radiation. Efficient heating takes place for converging axisymmetric cylindrical waves (under some limitations on the real part of the complex permittivity of the cylinder) if the diameter of the cylinder and the skin-depth are of the same order of magnitude and the electric field in the wave is directed along the common axis of the cylinder and the wave (see the exact conditions in Refs.~\cite{Akhm10,Akhm111}). This possibility can be used to create high energy density states for such applications as pumping of active media of short-wavelength lasers and nuclear fusion. For example, an exciting possibility of efficient heating of nanotubes by femtosecond laser pulses is discussed in Ref.~\cite{Akhm111}.

In this work we present the first experimental confirmation of the predictions of Refs.~\cite{Akhm4,Akhm10,Akhm111}) (some preliminary results for a somewhat problematic configuration were presented in Ref.~\cite{Akhm112}). In our experiment, a thin fiber (with a diameter three orders of magnitude smaller than the wavelength) absorbed up to 6\% of the microwave power focused on the fiber with an ellipsoidal reflector.

Work ~\cite{Akhm4,Akhm10,Akhm111} had important predecessors. Shevchenko (Ref.~\cite{Shev1}) derived optimal conditions of absorption of a plane electromagnetic wave in a thin conducting wire that are similar to the conditions of Refs.~\cite{Akhm4,Akhm10,Akhm111}. However, the possibility of efficient heating of a thin wire or fiber, when the power absorbed in a thin conducting wire or fiber is comparable to the power in the incident wave, was not noticed in work ~\cite{Shev1}, as heating by a plane wave is very inefficient. On the other hand, the transverse geometry of Refs.~\cite{Akhm4,Akhm10,Akhm111} (heating by a converging axisymmetric cylindrical wave) was also considered, e.g., in Ref.\cite{Zharov1}, but that work only contains resonant conditions of heating, which are difficult to use for practical plasma heating.

The results of Ref.~\cite{Shev1} found experimental confirmation in Ref.~\cite{Kuz1}.  The experimental results of Ref.~\cite{Kuz1}, motivated by the theoretical results for a plane wave, were obtained for absorption of microwave $\mathrm{H_{01}}$ mode at the output of a waveguide, and heating efficiency was not assessed. In the present experiment, we demonstrate efficient heating of a thin fiber by an electromagnetic beam in free space. The experimental results are in satisfactory agreement (typically up to a factor of 2) with theoretical computations.

\section{\label{sec:level1-2}The experimental setup}

\begin{figure*}
\includegraphics{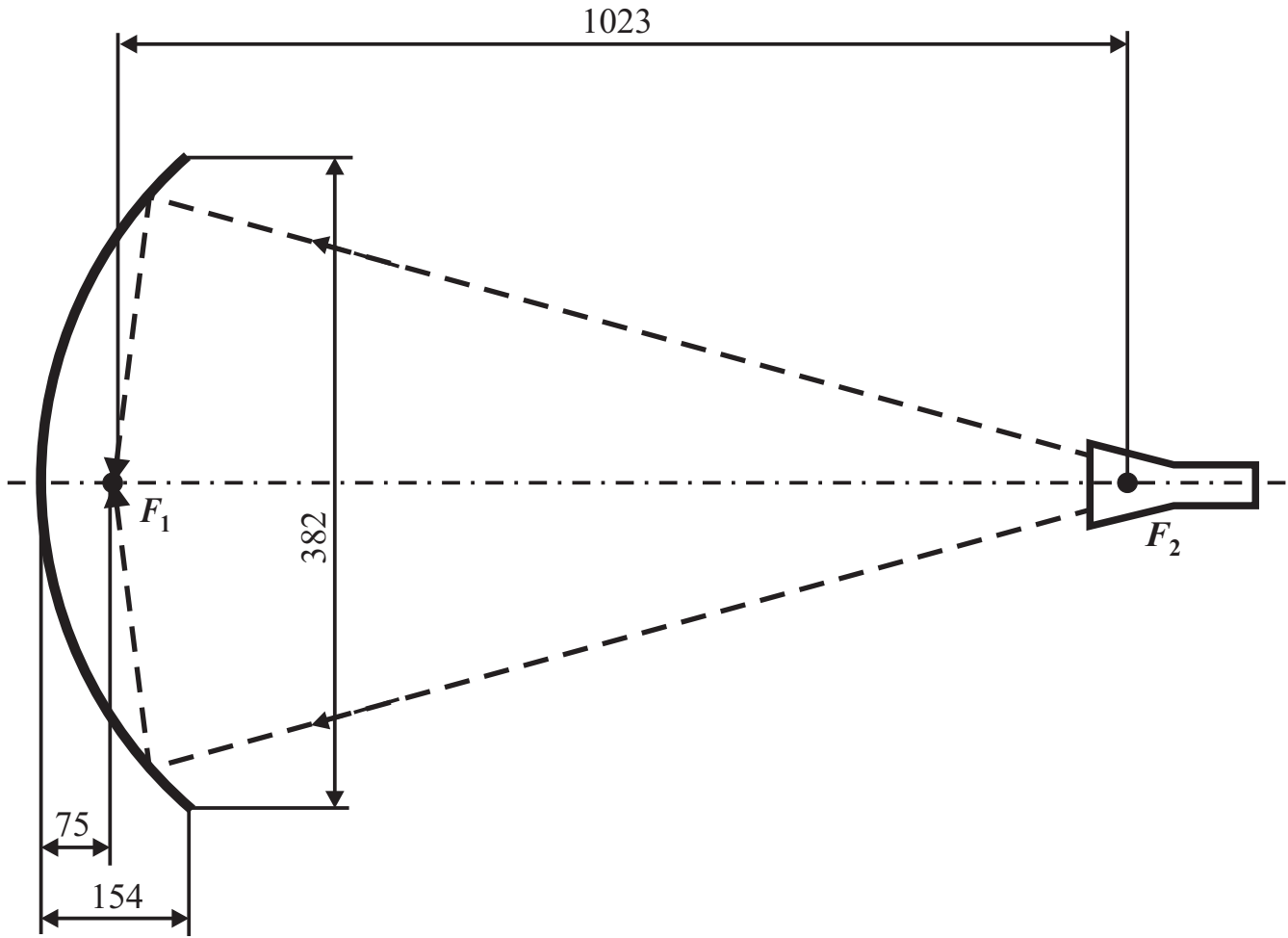}
\caption{\label{fig:fig1}The experimental setup (not to scale)}
\end{figure*}

The experimental setup is shown schematically in Fig.~\ref{fig:fig1}.
A thin wire or fiber is placed in focus $F_1$ of the ellipsoidal reflector. The internal surface of the reflector is a part of an ellipsoid of revolution defined by the equation:
\begin{equation}
\frac{x^2}{a^2}+\frac{y^2+z^2}{b^2}=1,\label{eq:1}
\end{equation}
where $a\approx$ 586 mm is the major semiaxis and $b\approx$ 287 mm is the minor semiaxis of the relevant ellipse. The distance between the foci $F_1$ and $F_2$ is approximately 1023 mm. The dimensions of the reflector and the position of focus $F_1$ inside the reflector are shown in Fig.~\ref{fig:fig1}. A horn with aperture 31 x 22 mm\textsuperscript{2} and length 130 mm is placed in focus $F_2$ (the distance between the horn aperture and the focus is 65 mm). The wide side of the horn is horizontal. The horn is connected to a waveguide with dimensions 7.2 x 3.4 mm\textsuperscript{2} with mode $H_{10}$. The frequency of the electromagnetic radiation varied from 24.5 GHz to 39 GHz, and the power varied from 20 mW to 61 mW. Typically, most of the power is collected by the reflector and focused in focus $F_1$.

\section{\label{sec:level1-3}Measurement and computation methods}
The wire or fiber in the focus is heated by the radiation, and the initial electrical resistance of the wire/fiber $R_0$  changes by $\Delta R$. The average wire/fiber temperature increase  $\Delta T$ corresponding to  $\Delta R$ was calculated as
\begin{equation}
\Delta T=\frac{\Delta R}{\alpha_r R_0},\label{eq:2}
\end{equation}
where $\alpha_r$ is the temperature coefficient of resistance.

On the other hand, the steady  state wire/fiber temperature increase depends on the absorbed power $P_a$ and the conditions of heat exchange with the environment \cite{Kuz1}:
\begin{equation}
\Delta T=\frac{P_a}{a_p L},\label{eq:3}
\end{equation}
where $L$ is the length of the wire/fiber and $a_p$ is the linear heat exchange coefficient. For the carbon fiber of diameter $11\mu$, this coefficient was measured as follows.
 For a direct current in the fiber, the Joule power was calculated based on the measurements of the current and the voltage, and the temperature increase
  was calculated based on the measured change of the fiber resistivity. The value of the coefficient differed slightly for a vertical and horizontal fiber:
  0.017 W/(m-deg) and 0.019 W/(m-deg), respectively. For the platinum wire of diameter $3.5\mu$, the following values obtained by N. G. Kokodiy and A. O. Pak (private communication) were used for the vertical and horizontal orientation of the wire: 0.023 W/(m-deg) and 0.026 W/(m-deg), respectively.

It follows from Eqs.~(\ref{eq:2},\ref{eq:3}) that
\begin{equation}
P_a=\frac{\alpha_p L}{\alpha_r}\frac{\Delta R}{R_0}.\label{eq:4}
\end{equation}
Therefore, the efficiency of absorption of microwave power in the wire/fiber equals:
\begin{equation}
K=\frac{P_a}{P}=\frac{\alpha_p L}{\alpha_r P}\frac{\Delta R}{R_0},\label{eq:5}
\end{equation}
where P is the power in the microwave beam.

In the experiment, microwave absorption in two targets was studied: 1) a platinum wire of diameter 3.5 micron and length $\approx$25 mm, and 2) a carbon fiber \cite{Cytec} of diameter 11 micron and length $\approx$30 mm. For the platinum wire, a tabular value $\alpha_r=$0.004 deg\textsuperscript{-1} was used, and for the carbon fiber, the value $\alpha_r=$-0.00021 deg\textsuperscript{-1} was found experimentally based on the current-voltage curve.

The experimental results were compared with the results of computations. The feed horn fields (incident on the reflector) were computed using the following formulas for the far zone of a pyramidal horn \cite{Balanis}.

If the origin is in the center of the horn aperture, and axes $x'$, $y'$, and $z'$ are directed parallel to the broad sides of the horn, parallel to the narrow sides of the horn, and along the horn axis, respectively, the radial distance, inclination angle, and azimuthal angle of the relevant spherical system of coordinates are $r$, $\theta$, and $\varphi$, respectively. The components of the magnetic field in the far zone are
\begin{eqnarray}
\nonumber
H_r(r,\theta,\varphi)=0,\\
\nonumber
H_\theta(r,\theta,\varphi)=-\frac{i k \exp(-i k r)}{4 \pi r}\frac{E_0}{\eta}I_1 I_2 \cos(\varphi)(1+\cos(\theta)),\\
H_\varphi(r,\theta,\varphi)=\frac{i k \exp(-i k r)}{4 \pi r}\frac{E_0}{\eta}I_1 I_2 \sin(\varphi)(1+\cos(\theta)),\label{eq:6}
\end{eqnarray}
where the temporal dependence factor $\exp(i \omega t)$ is omitted, $E_0$ is the amplitude of the electric field in the center of the horn aperture for the dominant mode,
\begin{widetext}
\begin{eqnarray}
\nonumber
I_1=\frac{1}{2}\sqrt{\frac{\pi\rho_2}{k}}\times\\
\nonumber
\left(\exp(i k'^{2}_x\rho_2/2 k)
\left(C(t'_2)-C(t'_1)-i\left(S(t'_2)-S(t'_1)\right)\right)+
\exp(i k''^{2}_x\rho_2/2 k)
\left(C(t''_2)-C(t''_1)-i\left(S(t''_2)-S(t''_1)\right)\right)\right) ,\\
I_2=\sqrt{\frac{\pi\rho_1}{k}}
\exp(i k^{2}_y\rho_2/2 k)\left(C(t_2)-C(t_1)-i\left(S(t_2)-S(t_1)\right)\right),\label{eq:7}
\end{eqnarray}
\end{widetext}
$C(x)$ and $S(x)$ are  the cosine and sine Fresnel integrals, respectively, $\rho_1$ is the distance from the horn aperture plane to the line of intersection of two opposite broad faces of the horn pyramid, $\rho_2$ is the distance from the horn aperture plane to the line of intersection of two opposite narrow faces of the horn pyramid,
\begin{eqnarray}
\nonumber
t'_1=\sqrt{\frac{1}{\pi k \rho_2}}\left(-\frac{k a_1}{2}-k'_x\rho_2\right),\\
\nonumber
t'_2=\sqrt{\frac{1}{\pi k \rho_2}}\left(+\frac{k a_1}{2}-k'_x\rho_2\right),\\
\nonumber
k'_x=k \sin(\theta)\cos(\varphi)+\frac{\pi}{a_1},\\
\nonumber
t''_1=\sqrt{\frac{1}{\pi k \rho_2}}\left(-\frac{k a_1}{2}-k''_x\rho_2\right),\\
\nonumber
t''_2=\sqrt{\frac{1}{\pi k \rho_2}}\left(+\frac{k a_1}{2}-k''_x\rho_2\right),\\
\nonumber
k''_x=k \sin(\theta)\cos(\varphi)-\frac{\pi}{a_1},\\
\nonumber
t_1=\sqrt{\frac{1}{\pi k \rho_1}}\left(-\frac{k b_1}{2}-k_y\rho_1\right),\\
\nonumber
t_2=\sqrt{\frac{1}{\pi k \rho_1}}\left(\frac{k b_1}{2}-k_y\rho_1\right),\\
\nonumber
k_y=k \sin(\theta)\sin(\varphi),\\
\end{eqnarray}
$a_1$ and $b_1$ are the lengths of the wide and the narrow sides of the horn aperture, respectively, and $\eta$ is the intrinsic impedance of the media (air).

The reflected fields for the ellipsoidal reflector were estimated using methods of physical optics \cite{Silver}, as the radii of curvature of the reflector are much greater than the wavelength everywhere. The reflected electric field $\boldsymbol{E}_s$ in a field point is calculated using the following formula (Ref.~\cite{Silver}):
\begin{eqnarray}
\nonumber
\boldsymbol{E}_s=\frac{1}{2\pi i \omega\varepsilon}\times\\
\int_{S_0}\left(\left(\boldsymbol{n}\times\boldsymbol{H}_i\right)\cdot\boldsymbol{\nabla}\left(\boldsymbol{\nabla}
\Psi\right)+k^2\left(\boldsymbol{n}\times\boldsymbol{H}_i\right)\Psi\right)dS,
\label{eq:9}
\end{eqnarray}
where $\Psi=\exp(-ikR/R)$, $R$ is the distance from the field point to the element of area $dS$ on the reflector, gradient operations in the integrand are referred to the field point as an origin, $\boldsymbol{n}$ is the normal to the reflector surface, $S_0$ is the geometrically illuminated surface of the reflector, components of the incident magnetic field $\boldsymbol{H}_i$ are defined by Eq.~(\ref{eq:6}), $\varepsilon$ is the media permittivity.

Absorption of the reflected field of the ellipsoidal reflector in the wire/fiber in the focus of the reflector was computed using the rigorous solution of the problem of diffraction of electromagnetic field on a homogeneous cylinder (Ref.~\cite{Wait1}), which has a simpler form in the case of axisymmetrical field (non-axisymmetrical field is not efficiently absorbed in a very thin cylinder). The field incident on the cylinder (wire/fiber) is described by the electric Hertz vector (Ref.~\cite{Stratton}) $\bm{\Pi}(\rho,\varphi,z)=\{0,0,\Pi(\rho,\varphi,z)\}$. We use a cylindrical coordinate system $\rho,\varphi,z$, where axis $z$ coincides with the axis of the cylinder. We do not use the magnetic Hertz vector as the relevant TE field is not efficiently absorbed in a thin cylinder. Therefore, the reflected field of the ellipsoidal reflector (which is the incident field for the cylinder) is defined by the following expansion of the $z$-component of the electric Hertz vector into cylindrical waves (Ref.~\cite{Stratton}):
\begin{eqnarray}
\Pi(\rho,\varphi,z)=\int d\gamma\alpha(\gamma)J_0(\lambda_1(\gamma)\rho)\exp(i \gamma z),\label{eq:1p2k}
\end{eqnarray}
where the limits of integration are $-\infty$ and $\infty$, $J_n(x)$ is the Bessel function of order $n$, $\lambda_1^2(\gamma)=\epsilon_1 k_0^2-\gamma^2$ (it is assumed that magnetic permeabilities of air and the cylinder, $\mu_1$ and $\mu_2$, equal 1), $\epsilon_1\approx1$ is the electric permittivity of air, $k_0=\omega/c$ is the wave vector in vacuum, $\omega=2\pi\nu$ is the frequency of the electromagnetic field (the factor $\exp(-\omega t)$ is omitted). Function $\alpha(\gamma)$ can be defined as follows. The $z$-component of the incident electric field corresponding to Eq.~(\ref{eq:1p2k}) can be written as follows:
\begin{eqnarray}
E_z(\rho,\varphi,z)=\int d\gamma\alpha(\gamma)\frac{\lambda_1^2(\gamma)}{\epsilon_1}J_0(\lambda_1(\gamma)\rho)\exp(i \gamma z),\label{eq:2p2k}
\end{eqnarray}
so
\begin{eqnarray}
\alpha(\gamma)\lambda_1^2(\gamma)=E_z(\gamma)=\frac{1}{2\pi}\int d z E_z(z)\exp(i\gamma z),\label{eq:3p2k}
\end{eqnarray}
where $E_z(z)$ is the $z$-component of the incident electric field on the axis of the cylinder (where $\rho=0$ and $J_0(\lambda_1(\gamma)\rho)=1$), computed using Eq.~(\ref{eq:9}), and $E_z(\gamma)$ is its Fourier transform. Eq.~(\ref{eq:2p2k}) correctly describes the $z$-component of the incident electric field in the vicinity of the axis, although Eq.~(\ref{eq:1p2k}) does not include the TE-field and non-axisymmetrical field, which are not efficiently absorbed in the thin cylinder.

The $z$-component of the electrical Hertz vector of the field refracted in the cylinder can be calculated using the rigorous solution of the problem of diffraction of electromagnetic field on a homogeneous cylinder (Ref.~\cite{Wait1,Akhm10}):
\begin{eqnarray}
u_2(\rho,\varphi,z)=\int d\gamma a_2(\gamma)J_0(\lambda_2(\gamma)\rho)\exp(i \gamma z),\label{eq:4p2k}
\end{eqnarray}
where
\begin{eqnarray}
a_2(\gamma)=\frac{\alpha(\gamma)\frac{\epsilon_2}{\epsilon_1}\frac{1}{J_0(p_2(\gamma))}\frac{-2 i}{\pi p_2^2(\gamma)H_0^{(1)}(p_1(\gamma))}}{-\left(\frac{1}{p_1(\gamma)}\frac{H_0^{(1)'}(p_1(\gamma))}
{H_0^{(1)}(p_1(\gamma))}-\frac{1}{p_2(\gamma)}\frac{\epsilon_2}{\epsilon_1}\frac{J_0'(p_2(\gamma))}
{J_0(p_2(\gamma))}\right)}=
\nonumber\\
=\frac{\alpha(\gamma)\epsilon_2\frac{1}{J_0(p_2(\gamma))}\frac{2 i}{\pi p_2^2(\gamma)H_0^{(1)}(p_1(\gamma))}}{\frac{1}{p_2(\gamma)}\epsilon_2\frac{J_1(p_2(\gamma))}
{J_0(p_2(\gamma))}-\frac{1}{p_1(\gamma)}\frac{H_1^{(1)}(p_1(\gamma))}
{H_0^{(1)}(p_1(\gamma))}},\label{eq:5p2k}
\end{eqnarray}
as $\epsilon_1 \approx 1$ and, for example, $H_0^{(1)'}=-H_1^{(1)}$, $\epsilon_2=\epsilon=\epsilon'+4\pi i \sigma/\omega$ is the complex electric permittivity of the cylinder, $\epsilon'$ is the real part of the permittivity, $\sigma$ is the conductivity of the cylinder, $p_1(\gamma)=\lambda_1(\gamma) a$, $p_2(\gamma)=\lambda_2(\gamma) a$, $a$ is the radius of the cylinder, $H_n^{(1)}(x)$ is the Hankel function, $\lambda_2^2(\gamma)=\epsilon_2 k_0^2-\gamma^2$.

The averaged $\rho$-component of the Poynting vector at the surface of the cylinder equals
\begin{eqnarray}
\frac{1}{2}\frac{c}{4\pi}\Re\left(\left(\bm{E}(a,\varphi,z)\times\bm{H}^*(a,\varphi,z)\right)_\rho\right)=
\nonumber\\
=-\frac{1}{2}\frac{c}{4\pi}\Re\left( E_z(a,\varphi,z)H_\varphi^*(a,\varphi,z)\right),\label{eq:6p2k}
\end{eqnarray}
as $H_z(a,\varphi,z)=0$ .

The total power absorbed in the cylinder $W$ equals
\begin{eqnarray}
-2\pi a \int d z \left(-\frac{1}{2}\frac{c}{4\pi}\right)\Re\left( E_z(a,\varphi,z)H_\varphi^*(a,\varphi,z)\right)=
\nonumber\\
=\frac{a c}{4 \pi} \int d z \Re\left( E_z(a,\varphi,z)H_\varphi^*(a,\varphi,z)\right)\label{eq:7p2k}
\end{eqnarray}
(an extra minus sign is introduced as positive $\rho$-component of the Poynting vector corresponds to energy flow out of the cylinder, and we are interested in the absorbed power.) Although the limits of integration are $-\infty$ and $\infty$, the reflected fields of the ellipsoidal reflector are negligible beyond the focal area. The components of the electric and magnetic field for the electric Hertz vector of Eq.~(\ref{eq:4p2k}) equal
\begin{widetext}
\begin{eqnarray}
E_z(a,\varphi,z)=\int d\gamma a_2(\gamma)\exp(i \gamma z)\frac{\lambda_2^2(\gamma)}{\epsilon_2}J_0(p_2(\gamma)),
\nonumber\\
H_\varphi^*(a,\varphi,z)=\int d\gamma' a_2^*(\gamma')\exp(-i \gamma' z)(-i k_0)\lambda_2^*(\gamma')J_0^{'*}(p_2(\gamma')),\label{eq:8p2k}
\end{eqnarray}
so
\begin{eqnarray}
\int d z E_z(a,\varphi,z)H_\varphi^*(a,\varphi,z)=
\nonumber\\
=\int d z \int d\gamma a_2(\gamma)\exp(i \gamma z)\frac{\lambda_2^2(\gamma)}{\epsilon_2}J_0(p_2(\gamma))\int d\gamma' a_2^*(\gamma')\exp(-i \gamma' z)(-i k_0)\lambda_2^*(\gamma')J_0^{'*}(p_2(\gamma'))=
\nonumber\\
=\int \int d\gamma d\gamma' a_2(\gamma)a_2^*(\gamma')\frac{\lambda_2^2(\gamma)}{\epsilon_2}(-i k_0)\lambda_2^*(\gamma')J_0(p_2(\gamma))J_0^{'*}(p_2(\gamma'))\int d z \exp(i \gamma z)\exp(-i \gamma' z)=
\nonumber\\
=\int \int d\gamma d\gamma' a_2(\gamma)a_2^*(\gamma')\frac{\lambda_2^2(\gamma)}{\epsilon_2}(-i k_0)\lambda_2^*(\gamma')J_0(p_2(\gamma))J_0^{'*}(p_2(\gamma'))2\pi\delta(\gamma-\gamma')=
\nonumber\\
=2 \pi \int d\gamma a_2(\gamma)a_2^*(\gamma)\lambda_2^2(\gamma)\lambda_2^*(\gamma)\frac{-i k_0}{\epsilon_2}J_0(p_2(\gamma))J_0^{'*}(p_2(\gamma)),\label{eq:9p2k}
\end{eqnarray}
\end{widetext}
and
\begin{eqnarray}
\Re\left(\frac{-i k_0}{\epsilon_2\lambda_2^*(\gamma)}J_0(p_2(\gamma))J_0^{'*}(p_2(\gamma))\right)=
\nonumber\\
=\Im\left(\frac{k_0}{\epsilon_2\lambda_2^*(\gamma)}J_0(p_2(\gamma))J_0^{'*}(p_2(\gamma))\right)=
\nonumber\\
=k_0 J_0(p_2(\gamma))J_0^{*}(p_2(\gamma))\Im\left(\frac{1}{\epsilon_2\lambda_2^*(\gamma)}
\frac{J_0^{'*}(p_2(\gamma))}{J_0^*(p_2(\gamma))}\right)=
\nonumber\\
=-k_0 J_0(p_2(\gamma))J_0^{*}(p_2(\gamma))\Im\left(\frac{1}{\epsilon_2^*\lambda_2(\gamma)}
\frac{J_0^{'}(p_2(\gamma))}{J_0(p_2(\gamma))}\right).\label{eq:10p2k}
\end{eqnarray}
Therefore, Eq.~(\ref{eq:7p2k}) can be rewritten as follows:
\begin{widetext}
\begin{eqnarray}
W=\frac{a c}{4}2 \pi\int d \gamma \left|a_2(\gamma)\lambda_2^2(\gamma)J_0(p_2(\gamma))\right|^2
(-k_0) \Im\left(\frac{1}{\epsilon_2^*\lambda_2(\gamma)}
\frac{J_0^{'}(p_2(\gamma))}{J_0(p_2(\gamma))}\right)=
\nonumber\\
=\frac{k_0 a c}{4}2 \pi\int d \gamma \left|a_2(\gamma)\lambda_2^2(\gamma)J_0(p_2(\gamma))\right|^2
\Im\left(\frac{1}{\epsilon_2^*\lambda_2(\gamma)}
\frac{J_1(p_2(\gamma))}{J_0(p_2(\gamma))}\right),\label{eq:11p2k}
\end{eqnarray}
\end{widetext}
as $J_0^{'}(x)=-J_1(x)$.

The radiated power of the pyramidal horn antenna (in the Gaussian system of units) equals \cite{Balanis}
\begin{eqnarray}
W_0=\frac{1}{4}\frac{c}{4\pi}a_1 b_1 E_0^2,\label{eq:12p2k}
\end{eqnarray}
where $a_1$ and $b_1$ are the dimensions of the horn aperture and $E_0$ is the amplitude of the electric field in the center of the aperture. Therefore, the heating efficiency (the part of the radiated power that is absorbed in the cylinder) equals
\begin{widetext}
\begin{eqnarray}
\frac{W}{W_0}=
\frac{8\pi^2 k_0 a\int d \gamma \left|a_2(\gamma)\lambda_2^2(\gamma)J_0(p_2(\gamma))\right|^2
\Im\left(\frac{1}{\epsilon_2^*\lambda_2(\gamma)}
\frac{J_1(p_2(\gamma))}{J_0(p_2(\gamma))}\right)}{a_1 b_1 E_0^2}.\label{eq:13p2k}
\end{eqnarray}
\end{widetext}

The following values of resistivity were used for the platinum wire and the carbon fiber, respectively: 0.106 $\mu$Ohm-m and 13 $\mu$Ohm-m.

While some manufacturer's data (Ref.~\cite{Cytec}) were used in computations for the carbon
 fiber, the experimental data for the specific fiber were somewhat different. For example, the fiber diameter was measured
  using diffraction of a broad laser beam on the fiber, and the measured value was $10.1\pm0.5$ micron, rather than 13 micron.
   The fiber resistivity was determined using measurements of the fiber resistance and dimensions. The value of resistivity
    was $16\pm2$ $\mu$Ohm-m, rather than 13 $\mu$Ohm-m. Using these parameters in computations did not result in significant modifications. For example, the part of power absorbed
     in the fiber changes from 9.7\% to 10.4\% at 39 GHz. The computed absorbed power changed insignificantly when 
     the value of the real part of the complex electric permittivity of the fiber changed, e.g., from -1 to 5.
\section{\label{sec:level1-3}Experimental and theoretical results}
\begin{figure*}
\includegraphics{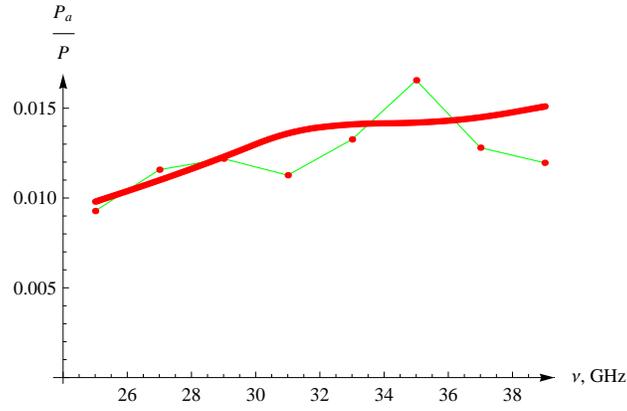}
\caption{\label{fig:processing2}Absorption of microwave radiation by a platinum wire in the focus of the reflector. Red line -- theoretical curve; green line -- experimental data.}
\end{figure*}

In Fig.~\ref{fig:processing2}, the dependence of absorption of electromagnetic power in the platinum wire on the frequency is shown. This case is not optimal for target heating, so only about 1\% of the beam power is absorbed in the wire.

The experimental values are in good agreement with the theoretical ones. The test point scattering is caused by environmental factors. The wire is heated just by 1$^\circ$ or less, so air flows can significantly influence the results.
\begin{figure*}
\includegraphics{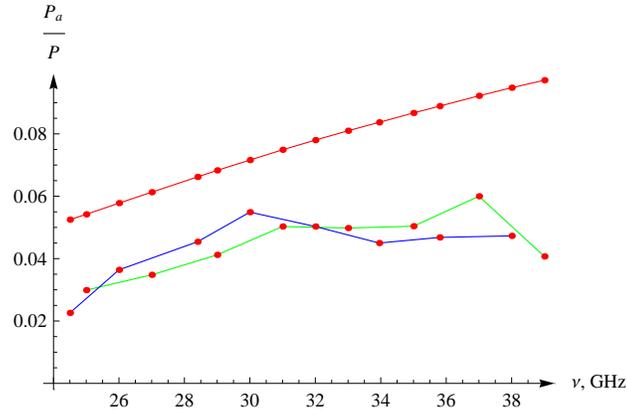}
\caption{\label{fig:processing1}Absorption of microwave radiation by a carbon fiber in the focus of the reflector. Red line -- theoretical curve; green and blue lines -- experimental data.}
\end{figure*}

In Fig.~\ref{fig:processing1}, the same dependence is shown for the carbon fiber. In this case, significantly more power is absorbed -- up to 6\%. The experimental values are less than the theoretical ones, but the agreement seems satisfactory. Two experimental curves are given in the figure. The blue line represents the results for the case where the reflector was covered by a sheet of polystyrene foam to decrease air flows. The foam refraction index is close to unity, so microwave reflection coefficient for normal incidence is of the order of 0.1\%. Therefore, the polystyrene foam sheet has very little effect
 on microwave propagation.

The discrepancy between theory and experiment may be due to inaccuracies of the fiber positioning or to a discrepancy between conductivity for direct current and for microwave frequencies.

To asses the effect of microwave field polarization on absorption in the fiber, the experiment was conducted for a horizontal orientation of the fiber, when the electric field is orthogonal to the wire. Within our approximations, the theoretical absorption efficiency is zero in this case. The experimental values
 (less than 0.4\%) were much less than for the other polarization, and the specific experimental values were not very reliable,
  as the fiber resistance changes were comparable to instrument error.

\begin{figure*}
\includegraphics{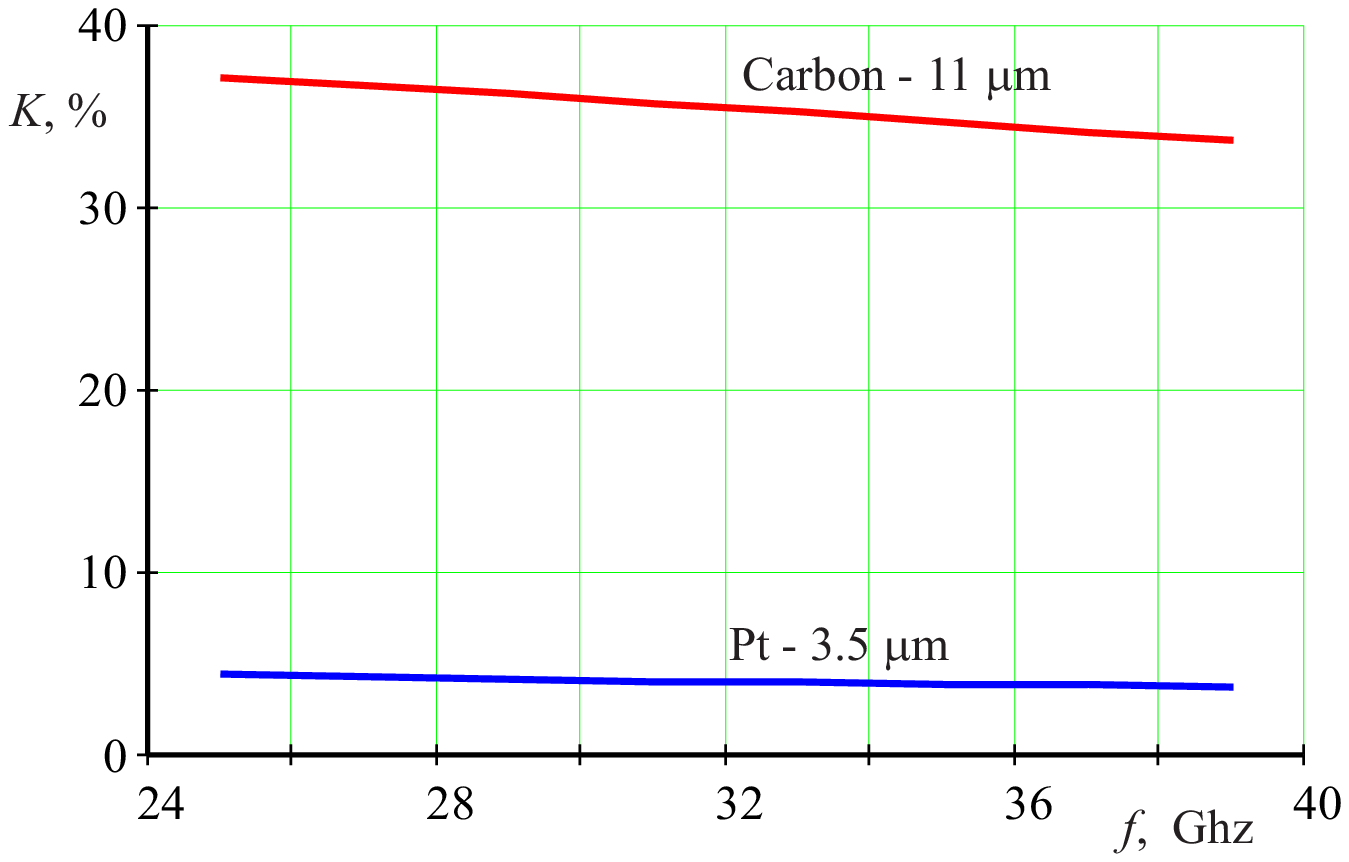}
\caption{\label{fig:fig4}Absorption of an axisymmetric converging cylindrical wave by a thin wire/fiber}
\end{figure*}

Absorption in the cylindrical target can be significantly greater for an axisymmetric converging cylindrical wave incident on the target. In this case, up to 40\% of the incident power can be absorbed in the carbon fiber (see Fig.~\ref{fig:fig4}).

The absorption in the platinum wire is significantly less -- about 4\%.

The increase of absorption with frequency in our experiment probably can be explained as follows: for higher frequency, the horn pattern is more narrow, so the reflector gathers more power.

In this experiment, the absorption is relatively modest -- about 6\% for the carbon fiber. This is due to the selected configuration. As continuous wave was used in the experiment, care was taken to exclude a possibility of multi-path heating of the target. Electromagnetic radiation reflected from the wire and then from the reflector would be directed to the other focus of the reflector. The absorption efficiency is proportional to the square of the angle from which the incident power is directed on the target. In our experiment, this angle is significantly less than 360$^\circ$ (about 200$^\circ$), so the efficiency is at least 3 times less than for the axisymmetrical cylindrical wave. However, in fast heating applications, the target can be irradiated from all directions, so higher efficiency can be achieved.

\section{\label{sec:level1-3}Conclusion}

The results of the experiment confirm the feasibility of efficient heating of thin cylindrical targets with electromagnetic radiation with wavelength (and, consequently, the dimensions of the focal area) several orders of magnitude greater than the diameter of the target. To this end, it is necessary to create an incident field with a high axisymmetric content (with respect to the axis of the target) and proper polarization, and there needs to be a match between the diameter of the target, its conductivity, and the wavelength. However, the conditions of efficient heating are non-resonant and therefore very promising for numerous applications. The heating efficiency of tens percent can be achieved for very thin targets.

\bibliography{dfcdb2art}

\end{document}